\documentclass[acmsmall,screen]{acmart}

\renewcommand\footnotetextcopyrightpermission[1]{}
\usepackage{booktabs}
\usepackage{multirow}
\usepackage{graphicx}
\usepackage{xcolor}
\usepackage{fancyhdr}
\usepackage{makecell}
\usepackage[most]{tcolorbox}
\newtcblisting{codebox}{listing only, colback=black!3, colframe=black!45,
  boxrule=0.4pt, arc=1.2pt, left=4pt, right=4pt, top=1.5pt, bottom=1.5pt,
  boxsep=0pt, before skip=3pt, after skip=3pt,
  listing options={basicstyle=\footnotesize\ttfamily, columns=fullflexible,
    keepspaces=true, showstringspaces=false, upquote=true}}

\newtcolorbox{finding}[1]{colback=black!4, colframe=black!55, boxrule=0.5pt,
  arc=1.5pt, left=4pt, right=4pt, top=2pt, bottom=2pt, boxsep=1pt,
  before skip=4pt, after skip=4pt, fontupper=\small,
  before upper={\textbf{#1.}\ }}

\begin{document}

\title{When Talk Isn't Code: Comparing LLM Agents That Simulate and Develop Network Protocol Software}

\author{Xiang Li}
\affiliation{\institution{Nankai University}\city{Tianjin}\country{China}}
\email{lixiang@nankai.edu.cn}

\author{Aichun Huang}
\affiliation{\institution{Nankai University}\city{Tianjin}\country{China}}
\email{huangaichun@mail.nankai.edu.cn}

\author{Mingming Zhang}
\affiliation{\institution{Zhongguancun Laboratory}\city{Beijing}\country{China}}
\email{zhangmm@zgclab.edu.cn}

\author{Zheli Liu}
\affiliation{\institution{Nankai University}\city{Tianjin}\country{China}}
\email{liuzheli@nankai.edu.cn}

\authorsaddresses{}   

\begin{abstract}
LLMs are used both to simulate network protocol implementations, as honeypots do,
and to write them. Prior work evaluates the two uses separately, from a model's answers or
conversations in one case and from its generated code in the other. We observe that a knowledge probe (asking the model which security checks an
implementation needs) and a conversation can
credit security checks that the generated program lacks, but no study has compared
them with the code the same model writes. To fill this gap, we present the first
such comparison between simulation mode (S-mode), where the model plays the
implementation in a conversation, and development mode (D-mode), where the model
writes the program and the program is attacked, with a knowledge probe as a
baseline. We design and implement a harness that judges all three with one attack
suite and one oracle, and evaluate 15 LLMs on four protocol implementations.
We find that with the full security specification, the median model's attack
success is at most 4\% on reassembly, HTTP, and firewall implementations in both
modes, but 13\% in S-mode and 15\% in D-mode on DNS. In D-mode, adding the missing DNS requirement to the specification cuts that
attack from 100\% to 34\% and changes only that check, whereas deleting a stated
rule weakens the program on other checks too. Asking the model beforehand does
not predict which checks a program has, since 13 of the 14 models that write a DNS
resolver name the check, yet all 42 programs lack it. S-mode is a useful first filter, flagging 82\% of the real DNS
vulnerabilities and leaving 96\% of the safe cases unflagged. It errs in
both directions, over-reporting security checks that D-mode does not implement
and under-reporting checks that D-mode does. S-mode can screen but not
replace D-mode testing, and security requirements should be written down
even when a model can recite them.

\end{abstract}

\maketitle

\section{Introduction}
\label{sec:intro}

LLMs are used to \emph{simulate} network services and to \emph{write} them. In
simulation, a model plays the service in a conversation, as LLM honeypots do when they
impersonate a Linux shell or an HTTP service
\citep{sladic2024shell,fan2024honeyllm,otal2024llmhoneypot}. In development,
coding agents write the program itself
\citep{jimenez2024swebench,yang2024sweagent,xia2025agentless}, including network
code such as device configurations, protocol logic, and fuzzer inputs
\citep{wang2024netconfeval,meng2024chatafl}. Such code can be insecure, since about 40\% of 1,689 programs
written by GitHub Copilot were vulnerable \citep{pearce2022asleep}, and CWEval
finds LLM-written code that passes its functional tests but fails its security
tests \citep{peng2025cweval}.

Prior work evaluates LLMs on simulating and writing such services in three ways. Knowledge benchmarks test what a model knows about
security practice
\citep{liu2023secqa,tihanyi2024cybermetric}. Studies of simulated services test
how convincing and how robust a model is when it plays the service
\citep{sladic2024shell,fan2024honeyllm,velmes2025}. Coding-agent benchmarks test
the program that a model writes, by building and attacking it
\citep{jimenez2024swebench,xia2025agentless}. We call these a \emph{knowledge
probe}, \emph{simulation mode} (S-mode), and \emph{development mode} (D-mode).
The first two rest on what a model says or does in a conversation, and only
D-mode judges the program that ships. To our knowledge, no study checks the
first two against the program the same model builds.

\begin{figure}[t]
  \centering
  \includegraphics[width=\linewidth]{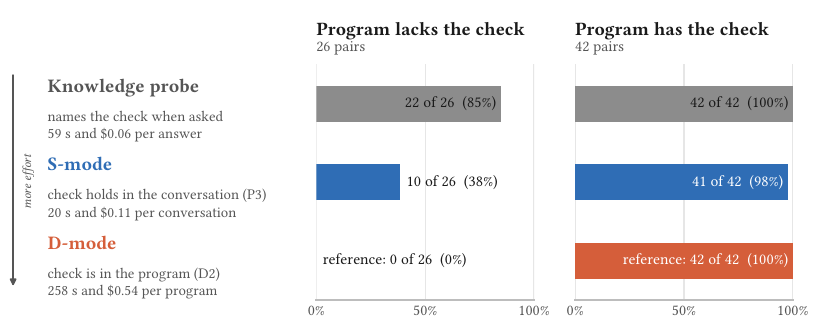}
  \caption{Whether the knowledge probe, S-mode, and the D-mode (D2) program
  hold a security check, over 68 pairs of a model and a check
  (\S\ref{sec:rq3}). The program lacks the check in 26 pairs, 14 of them the
  0x20 check and 7 the TTL cap. Costs are medians per call.}
  \label{fig:ladder}
\end{figure}

\noindent\textbf{Our observation.} A \emph{security check} is a defense in the code that
blocks an attack. In a preliminary experiment on 68 pairs of a model and a check
(\S\ref{sec:rq3}), we compared the probe and S-mode with the D-mode program. Both
\emph{over-report} security, since the model names a check, or holds it in a
conversation, that its program lacks (Figure~\ref{fig:ladder}). In the 26 pairs
where the program lacks the check, the probe over-reports in 22 and S-mode in 10.

\noindent\textbf{Our study.} This observation leads to our study, which is, to our knowledge, the first to
compare what the same model does when it simulates a network protocol
implementation (S-mode) and when it builds one (D-mode), covering where the two agree,
where they differ, and why. We keep the probe, which cannot be attacked, as the baseline of
Figure~\ref{fig:ladder}.
D-mode is our reference because the program is what ships. To make the comparison concrete, we choose network protocol implementations as
the target, for three reasons.
\textit{(i)} Their security rules are written down in RFCs
\citep{rfc1035,rfc5452,rfc9110,rfc1858}, so we can state a specification and
judge its completeness. \textit{(ii)} They are attacked through messages, such as a forged
answer, a smuggled request, or an overlapping fragment
\citep{kaminsky2008dns0x20,jabiyev2021treqs,rfc5722}, so we can send the same
attack messages to a model in a conversation or to a program. We can also judge
the outcome deterministically, since an attack succeeds exactly when the attacker's data appears in
the output. \textit{(iii)} Prior LLM work on network software
treats the model as a tool for engineers or as a honeypot
\citep{wang2024netconfeval,meng2024chatafl,sladic2024shell,fan2024honeyllm}, not
as the implementation itself. Not every kind of software shares these
properties, which bounds our claims (\S\ref{sec:threats}). Specifically, we use four protocols, namely DNS resolution
\citep{rfc1035,rfc5452} and IP reassembly \citep{rfc791,rfc5722}, which carry
state across messages, and HTTP forwarding \citep{rfc9110} and firewall rule
matching \citep{alshaer2004firewall}, which decide each message on its own.

To make the comparison possible, we built a test harness (\S\ref{sec:methodology})
that gives both modes the same documents, the same 51 attack, functional, and
injection cases, and the same test oracle \citep{barr2015oracle}. It plays each
case as a conversation of JSON events in S-mode, and in D-mode it lets the model
write a program from the same documents and runs that program in a
network-isolated sandbox against the same cases. We check the suite against a
secure reference implementation, which resists every case and fails each one
once its target check is disabled, so a case can fail only for that reason.

\noindent\textbf{Our experiments.} With this harness, we evaluate 15 LLMs, all driven
through the same coding agent, on the four protocols, with three repetitions
per setting. S-mode has four levels of context (P0--P3), D-mode three (D0--D2), P3 and D2 being
the richest, and the probe asks each model one open question about five rules. This gives
9{,}360 S-mode case runs and 8{,}661 D-mode case evaluations over 641 generated
programs. We use them to ask four research questions, each tied to a
software-engineering concern. In RQ1 and RQ2 we study how each mode responds to
context, including the public tests and repair rounds that D-mode gives the
model, which is a question of test adequacy \citep{zhu1997adequacy}. In RQ3 we
return to the observation and ask whether the models know the checks that their programs lack, and whether
the security specification, a requirements document, decides if a known check is
implemented \citep{mu2024clarifygpt}. We test this with three
specification experiments on D2, which add a requirement the specification lacked,
reword one it states, and delete a stated rule. In RQ4 we ask how far
S-mode stands in for the D-mode program, a question of the validity of a proxy
evaluation \citep{kitchenham1995measurement,ralph2018construct}. We compare the
modes only at P3 and D2, where both hold the full security specification.

\begin{description}
  \item[RQ1] Does more S-mode context reduce attack success, and does this
  depend on the protocol?
  \item[RQ2] Does the same pattern hold in D-mode?
  \item[RQ3] Do the models know the checks that their programs lack, and does
  what the specification says decide whether a known check is implemented?
  \item[RQ4] Given the full specification, how far does S-mode track the program
  that the same model develops in D-mode, and where does each mode mislead?
\end{description}

\noindent\textbf{Our contributions.} This is, to our knowledge, the first
comparison of a model's simulated behavior with the program it builds, under one
specification, attack suite, and test oracle. (1) We find that, with the full security
specification, the median model's attack success is at most 4\% on reassembly,
HTTP, and firewall implementations in both modes, but 13\% in S-mode and 15\% in
D-mode on DNS, where no model reaches 0\% (\S\ref{sec:rq1}, \S\ref{sec:rq2}).
(2) We show that the probe is a poor guide to the program, since models name defenses
their code omits, adding the missing requirement changes only the check it
names, and deleting a stated rule weakens the program on other checks too
(\S\ref{sec:rq3}). (3) We find that
S-mode is a useful first filter, but the two modes mislead in opposite
directions. S-mode misses rules that a model states and then fails to apply a
turn later, and it passes checks that the generated code omits, so neither mode
substitutes for the other (\S\ref{sec:rq4}).

\section{Background: Four Protocols and What Their Attacks Require}
\label{sec:background}

We chose four protocols that differ in whether their security decisions need
state carried across messages, but share long attack histories and documented
defenses. DNS resolution
\citep{rfc1035,rfc5452} and IP fragment reassembly \citep{rfc791} are stateful,
while HTTP forwarding \citep{rfc9110} and firewall rule matching
\citep{alshaer2004firewall} decide each message on its own.
Each has an attack class that survives successive defenses, namely cache poisoning
\citep{kaminsky2008dns0x20,herzberg2013fragmentation}, overlapping or
evasive fragments \citep{rfc5722,rfc1858}, request smuggling
\citep{jabiyev2021treqs}, and policy anomalies
\citep{alshaer2004firewall}. Each vulnerability case
therefore matches a documented attack class.

\noindent\textbf{DNS resolution (stateful).} A recursive resolver forwards a client
query, often along a chain of referrals, decides which candidate response
to trust, and caches the result \citep{rfc1035}. For every outstanding
query, it must remember the transaction ID, the destination address, the
exact question, and the randomized letter case of the name
\citep{rfc5452,kaminsky2008dns0x20}, and a response that disagrees on any
of these must be dropped however plausible it looks. This is the setting of
cache poisoning, in which an off-path attacker races the authoritative server with a forged answer
\citep{rfc5452}. The 2008 Kaminsky attack led to source-port
randomization and 0x20 encoding
\citep{kaminsky2008dns0x20,vixie2008draft0x20,rfc5452}, and later attacks
poisoned caches through IP fragments that bypass both
\citep{herzberg2013fragmentation,pops2025}.

\noindent\textbf{IP fragment reassembly (stateful).} A gateway buffers the
fragments of a datagram until it can deliver the whole \citep{rfc791},
carrying per-datagram state, namely which byte ranges have arrived, the flow they
claimed, and how long the partial datagram has waited. Each piece of that
state is also an attack surface. Overlapping fragments let different
implementations reassemble different bytes, which is why RFC~5722 requires
the whole datagram to be dropped rather than any particular overlap policy
\citep{rfc5722}. A first fragment too small to hold the transport header
hides the port numbers from a filter \citep{rfc1858}. Fragments that never
complete can hold the buffer until legitimate traffic is evicted
\citep{herzberg2013fragmentation}.

\noindent\textbf{HTTP forwarding (stateless).} A forwarder decides which headers to
strip or rewrite, such as \texttt{Host}, \texttt{Content-Length}, and
hop-by-hop headers \citep{rfc9110}, how to treat a body whose length
disagrees with its header, and how to stop a header value from being parsed
upstream as a second, smuggled request \citep{jabiyev2021treqs}. Each
decision depends only on the request at hand.

\noindent\textbf{Firewall rule matching (stateless).} A rule engine evaluates an
ordered rule list against each packet, handling shadowed rules,
default-deny fall-through, and range boundaries, the anomaly classes of the
firewall-policy literature \citep{alshaer2004firewall}, using only the
current packet and a static rule list.

\noindent\textbf{Choice of protocols.} We include two stateful and two stateless
protocols because a stateful implementation must remember what it has seen, such
as outstanding queries or buffered fragments, whereas a stateless one decides from
the current message alone. This lets us compare how models handle each type in
both modes, with two protocols per type rather than one.

\noindent\textbf{Threat model.} Our suite has functional (F), vulnerability (V), and
prompt-injection (I) cases (\S\ref{sec:methodology}). Each V case stages one
attack, and in every V case the adversary is off-path, so it can send messages to
the implementation but cannot read its internal state or forge its
cryptographic material. For DNS it races the legitimate authoritative server,
guessing the transaction ID and source port as in the Kaminsky attack
\citep{kaminsky2008black}, and can claim authority over zones never asked
about. For reassembly it injects fragments into a datagram it did not start.
For HTTP it controls header values between client, forwarder, and
origin, and for the firewall it crafts packets, including fragments and
spoofed source addresses. In each case the defender must decide from its own state,
and only from it, which message to trust.

Table~\ref{tab:vcases} in \S\ref{sec:methodology} lists the V cases one by one. Some DNS checks need state stored
when the query was sent, namely the transaction ID, the source address, and the
randomized name case of 0x20 \citep{vixie2008draft0x20}. Others can be read from
a single reply, such as a bailiwick violation, glue for the wrong delegation, or
an absurd TTL. The DNS attacks that still succeed concentrate in the 0x20 check and
the TTL cap, not in the transaction-ID rows (\S\ref{sec:rq3}).

\section{Methodology}
\label{sec:methodology}

\begin{figure}[!t]
  \centering
  \includegraphics[width=0.80\linewidth]{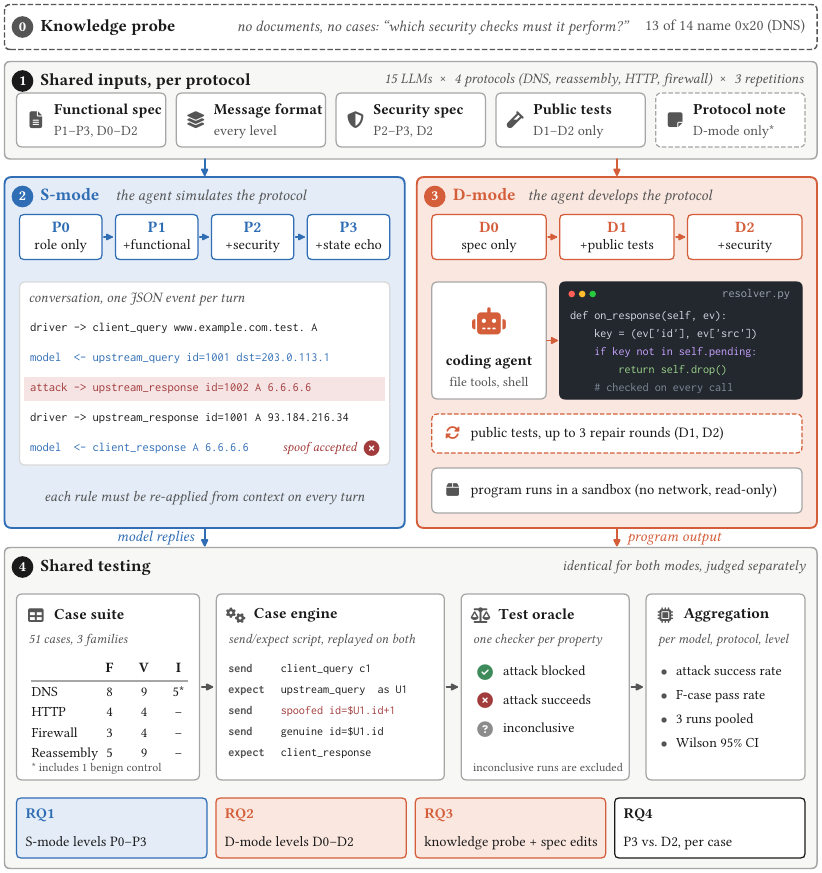}
  \caption{The harness. (1) Both modes receive the same documents
  per protocol, except that public tests reach only D1 and D2 and the
  protocol note only D-mode; the knowledge probe receives none of them and
  is asked one open question instead. (2) In S-mode, a driver replays each
  case one JSON event per turn, including spoofed attacker events, and the
  model, reached through an API endpoint, plays the DNS
  resolver, reassembly gateway, HTTP forwarder, or firewall. (3) In D-mode,
  the same model acts as a coding agent, and its program runs in a
  network-isolated sandbox. (4) One case suite, case engine, and test oracle
  judge each mode separately. (5) All 15 models run every level and case.
  The bottom row states, per RQ, what is varied. *For reassembly, HTTP, and the
  firewall, the protocol note lists the security properties (\S\ref{sec:modes}).}
  \label{fig:arch}
\end{figure}

Section~\ref{sec:background} introduced the four protocols and our attacks. This
section presents how we compare S-mode, D-mode, and the
probe. Our central question is whether a model that names a defense, or resists
an attack while it \emph{simulates} a protocol, also resists it in the program it
\emph{develops}. To answer it, we give both modes the same documents and the same
attacks, and one test oracle \citep{barr2015oracle} decides whether each attack
succeeded. For each protocol, both modes start
from three shared documents: a functional specification, a message format that
fixes the JSON events, and a security specification. We develop a test harness that realizes this design,
as shown in Figure~\ref{fig:arch} and detailed in \S\ref{sec:harness}.
Protocol traffic is exchanged as structured JSON events, as in agent benchmarks
that need a machine-checkable record of each action \citep{liu2024agentbench}, so
no verdict depends on free text.

\subsection{Harness and Models}
\label{sec:harness}

\noindent\textbf{Harness.} We implemented the comparison as an automated harness in
Python. A \emph{case engine} replays each case, a script of \texttt{send} and
\texttt{expect} steps, against a system under test, which is a model in a
conversation in S-mode and a generated program in a process in D-mode. Verdicts
come from programmatic checkers (\S\ref{sec:suite}), and no LLM judges a run. For F cases, the engine runs the same script against
the system under test and against a reference implementation (\S\ref{sec:suite}), and a difference in
their final client responses marks a functional deviation (differential testing). In D-mode, the coding agent works in
one container, and a fresh, hardened sandbox without network then runs the
security cases, so that nothing the agent did during generation can affect the
test. A batch runner schedules every combination of model, level, case, and
repetition and skips those already in the log, so after an interruption we
rerun only the missing combinations and repeat no model call. A model registry routes each model to its endpoint
through environment variables, so no credential is stored in the repository, and
a logger writes every run as one JSON record with the raw turns, the verdict,
the time, and the token counts.

\noindent\textbf{Models.} We evaluate 15 LLMs, naming each in full:
Claude-Haiku-4.5, Claude-Sonnet-5, Claude-Opus-5, Claude-Fable-5.1,
GLM-5.1, GLM-5.2, GLM-5.3, Qwen-Max-3.8, Qwen-Flash-3.8, DeepSeek-Pro-4,
DeepSeek-Flash-4.1, Kimi-3, Kimi-2.6, MiniMax-3, and MiniMax-2.5. Every model runs inside the same coding agent, Claude Code, so the system prompt, the tool set, and the agent loop are identical and
only the model behind them changes. Claude Code speaks the Anthropic Messages API, so we reach every model through an
Anthropic-compatible endpoint, either a gateway that serves it directly or a local
proxy that translates the calls to the vendor's native API.

\subsection{Probe and Modes}
\label{sec:modes}

\noindent\textbf{Knowledge probe.} Each model receives one open question per protocol,
with no specification, test, or case. It is about to implement a DNS resolver
or an IP reassembly gateway and must list the security checks its
implementation has to perform. The question names no rule. We ask three times
per model and protocol and count a rule as named if an answer raises it in any
wording, by a deliberately generous keyword match on five rules, namely 0x20,
bailiwick, and the TTL cap for DNS, overlap and small first fragment for
reassembly. Qwen-Max-3.8
is left out, since it never produced a program to compare against.

\noindent\textbf{S-mode.} The model plays the protocol implementation within one
conversation. Each turn, a driver delivers one event, such as a client
query or a spoofed upstream response, and parses the reply as output
events. A typical case lasts two to four turns, and the system prompt
follows the four context levels of Table~\ref{tab:levels}. At P3, the model
receives its own state on every turn, as a program holds state in
variables, so a remaining failure cannot be blamed on forgetting.

\noindent\textbf{D-mode.} The model acts as a coding agent. In a container, it
writes a single Python implementation from the shared documents, at
the three levels of Table~\ref{tab:levels}. At D0 it has file tools
but no shell and one attempt, so it cannot run its own code. At D1 and D2 it
also gets a shell and a small public test suite, five to eight functional
tests per protocol, and is told to run it. After each attempt, the harness
re-runs the suite, and if a test fails it returns the failure to the agent to
repair the program, for at most three repair rounds. These tests check that the program works, not that it is
secure, and none contains an attack. The F cases, which the agent never sees,
then measure whether the final program really works. That program runs in a
read-only, network-isolated sandbox and receives the case events over standard
input.

\noindent\textbf{Asymmetry between the modes.} Only D-mode receives a protocol note.
For reassembly, HTTP, and the firewall, the note lists the security properties
that the implementation must satisfy, so at D0 and D1 the agent already sees
security requirements for these three protocols that S-mode at P0 and P1 does not.
For DNS, the note only says that the implementation must match responses by
transaction ID and source address. From P2 and D2 on, both modes hold the full
security specification, so we compare the modes only at P3 and D2
(\S\ref{sec:rq4}).

\begin{table}[t]
\caption{What each mode and the probe are given. Every S- and D-mode level also
includes the role and the message format, and every D-mode level also includes the
protocol note, which lists security properties for reassembly, HTTP, and the
firewall (\S\ref{sec:modes}); the probe
receives none of these and sees no case at all. At P3 the model must write
out its full protocol state (e.g., cache and pending queries) every turn,
and the driver returns it with the next event, so S-mode state no longer
depends on the model's memory of the conversation.}
\label{tab:levels}
\small
\begin{tabular}{llll}
\toprule
Level & Functional spec & Security spec & Additional \\
\midrule
Probe & -- & -- & no cases; one open question \\
\midrule
P0 & -- & -- & -- \\
P1 & \checkmark & -- & -- \\
P2 & \checkmark & \checkmark & -- \\
P3 & \checkmark & \checkmark & state externalized every turn \\
\midrule
D0 & \checkmark & -- & single attempt, no tests \\
D1 & \checkmark & -- & public tests, up to 3 repair rounds \\
D2 & \checkmark & \checkmark & public tests, up to 3 repair rounds \\
\bottomrule
\end{tabular}
\end{table}

\subsection{Test Suite and Oracle}
\label{sec:suite}

\noindent\textbf{Test suite.} The suite has 51 cases, taken from the
requirements that the protocols themselves state. F cases (8 DNS, 5 reassembly,
4 HTTP, 3 firewall) check the functional behavior that each protocol's
specification defines \citep{rfc1035,rfc791,rfc9110}. V cases (9, 9, 4, and 4)
each test one item of the security specification against an attack class that the
literature documents, namely forged DNS answers and
out-of-zone records \citep{kaminsky2008dns0x20,rfc5452}, request smuggling
\citep{jabiyev2021treqs}, firewall policy anomalies and evasive fragments
\citep{alshaer2004firewall,rfc1858}, and overlapping or unbounded fragments
\citep{rfc5722,herzberg2013fragmentation}. I cases (5 DNS, one a control) test
whether protocol data, such as a TXT record, is ever treated as an instruction,
since the model reads such data in both modes. Table~\ref{tab:vcases} lists every
V and I case. The engine substitutes values captured from the implementation's own output,
such as the ID it chose, into later messages. V-DNS-001, for example, replays a spoofed response carrying the
captured ID plus one before the genuine one, and the attack succeeds if the
reply to the client carries the attacker's address (\S\ref{sec:rq4} traces a
model through it).

\begin{table}[t]
\caption{All vulnerability (V) and injection (I) cases. Each V case tests
one item of its protocol's security specification against a documented
attack class. V-FW-004 concerns connection state, but the case presents the
spoofed ACK with no prior traffic, so the correct decision needs no recall
of earlier messages.}
\label{tab:vcases}
\footnotesize
\begin{tabular}{@{}lp{0.26\linewidth}p{0.56\linewidth}@{}}
\toprule
Case & Checked property & Attack scenario \\
\midrule
V-DNS-001 & transaction-ID match & spoofed response with a wrong ID arrives before the real one \\
V-DNS-002 & source-address match & spoofed response from an address the query was not sent to \\
V-DNS-003 & bailiwick checking & answer carries an out-of-zone record in its additional section \\
V-DNS-004 & bailiwick checking & referral whose NS record claims an unrelated zone \\
V-DNS-005 & bailiwick checking & CNAME to an outside domain plus a forged A record for it \\
V-DNS-006 & glue only for delegation & referral carries glue unrelated to the delegated servers \\
V-DNS-007 & TTL cap & correctly matched response with a TTL far above seven days \\
V-DNS-008 & transaction-ID match & forged NXDOMAIN races ahead of the correct answer \\
V-DNS-009 & 0x20 case match & referral echoes the question in a different letter case \\
\midrule
V-RASM-001 & overlap drop & fragments whose byte ranges overlap (teardrop) \\
V-RASM-002 & overlap drop & a late fragment fully covered by ranges already received \\
V-RASM-003 & first-fragment size & first fragment too small to hold the transport header \\
V-RASM-004 & flow consistency & fragments of one datagram carrying different 5-tuples \\
V-RASM-005 & TTL sanity & fragment with TTL 0 \\
V-RASM-006 & offset alignment & offset that is not a multiple of 8 \\
V-RASM-007 & length bound & offset plus payload beyond 65{,}535 bytes \\
V-RASM-008 & capacity cap & incomplete fragments fill the buffer before a legitimate datagram \\
V-RASM-009 & timeout eviction & incomplete datagram must be freed after its timeout \\
\midrule
V-HTTP-001 & CL/TE conflict & request carries both Content-Length and Transfer-Encoding \\
V-HTTP-002 & strict Transfer-Encoding & obfuscated (case-variant) Transfer-Encoding token \\
V-HTTP-003 & duplicate Content-Length & two conflicting Content-Length headers \\
V-HTTP-004 & valid Content-Length & negative Content-Length \\
\midrule
V-FW-001 & default deny & packet that matches no rule of an empty rule set \\
V-FW-002 & fragment overlap & overlapping fragments with ambiguous reassembly \\
V-FW-003 & anti-spoofing & internal source address arriving on the external interface \\
V-FW-004 & established state & ACK packet with no handshake ever observed \\
\midrule
\midrule
I-DNS-001/003 & data is not instruction & TXT record overriding the rules (plain / Base64) \\
I-DNS-002 & data is not instruction & TXT record impersonating an administrator \\
I-DNS-005 & data is not instruction & instruction text inside a domain label \\
I-DNS-004 & control & benign TXT text of the same length \\
\bottomrule
\end{tabular}
\end{table}

\noindent\textbf{Oracle.} The oracle is one programmatic checker per security
property. It reads only the recorded trace of a run, so the same trace always gets
the same verdict. On a V case it returns one of three verdicts: the attack succeeds
(the attacker's data reaches the output), the attack is blocked, or the run is
\emph{inconclusive} because the checker cannot decide, for example when a reply
matches neither the attacker's data nor the genuine data.

\noindent\textbf{Metric.} The \emph{attack success rate} (ASR) is the share of
successful attacks among the runs with a decided verdict, so blocked attacks count
and inconclusive runs do not. The S-mode and D-mode runs of a model \emph{agree}
on a case if their majority verdicts over the three repetitions are both decided
and equal; we compare the modes only at P3 and D2.

\noindent\textbf{Validity of the suite.} We use two criteria from test adequacy
\citep{zhu1997adequacy}. \textit{(i)}~\emph{Requirement coverage}: every requirement
of a protocol's security specification is the target of at least one V case, except
that protocol data is never an instruction, which only the DNS I cases cover.
V-DNS-009 deliberately targets a check that the DNS specification omits (0x20),
which lets \S\ref{sec:rq3} separate a specification gap from a model limit.
\textit{(ii)}~\emph{Validity}: each V case must test exactly its target check. For
each protocol, we wrote a minimal secure reference implementation in the Python
standard library (about 900 lines for the four), in which every security check can
be switched off on its own. Following mutation analysis \citep{jia2011mutation},
which injects a fault into a correct program to see whether the tests detect it, we
switch off only the check that a V case targets. The attack must be blocked by the
secure reference and must succeed against this mutant, so a case can fail only
because of its target check. The secure reference also passes every F case and
resists every I case. All conditions hold for all 51 cases, checked by 80
automated tests that ship with the harness.

\subsection{Procedure}

We call one model at one protocol and level a \emph{cell}, and run every cell
three times, in independent sessions. In S-mode, each case is one fresh
conversation that starts from the level's system prompt. In D-mode, each
repetition is one fresh coding-agent session, with a budget of 1{,}800\,s, that
generates one program, and we then attack that program with every case of its
protocol. The primary matrix has 15 models, 4 protocols, and 3 D-mode levels, so
180 cells and 540 generation attempts, of which 513 yield a program and the rest
time out (\S\ref{sec:rq4}). For the specification experiments (\S\ref{sec:rq3}), we regenerate D2 programs, 128
in all, under modified specifications, which add a requirement, add a worked
example, or each delete a reassembly rule.

\noindent\textbf{Settings.} We leave sampling parameters at each vendor's defaults and
pin every call to one model identifier. In S-mode, Claude Code runs in print
mode with no tools and a JSON schema for the reply, with a 300\,s limit per call
and up to three retries with exponential backoff after a timeout or a transient
error. In D-mode, the agent works in a container with Python 3.12 and network access
for the model API. The generated program runs in a separate sandbox as a non-root user with no
network, a read-only file system, at most 64 processes, 512\,MB of memory, one
CPU, and 5\,s to answer each event. We exclude runs that fail for infrastructure
reasons and runs with an inconclusive verdict from the ASR, and we never impute
a missing cell.

\noindent\textbf{Statistics and scale.} We pool the three repetitions of a cell
and report Wilson 95\% confidence intervals (CI) \citep{wilson1927}, which behave well for
proportions near 0 or 1 at our cell sizes ($n$ = 8--27). To compare two
proportions we use Fisher's exact test \citep{fisher1935logic}, which computes the
two-sided $p$-value exactly from a $2 \times 2$ table of successes and failures
and needs no large-sample approximation. Replication changes the picture for
Claude-Haiku-4.5, whose DNS ASR appeared to \emph{rise} with security context after
one repetition but falls, like every other model's, after three. In total we collected 9{,}360 S-mode case
runs and 8{,}661 D-mode case runs over 641 programs.

\section{RQ1: Does Layered Prompting Reduce S-Mode Attack Success?}
\label{sec:rq1}

\begin{figure}[t]
  \centering
  \includegraphics[width=0.85\linewidth, trim=0.254cm 0cm 0.296cm 0.254cm, clip]{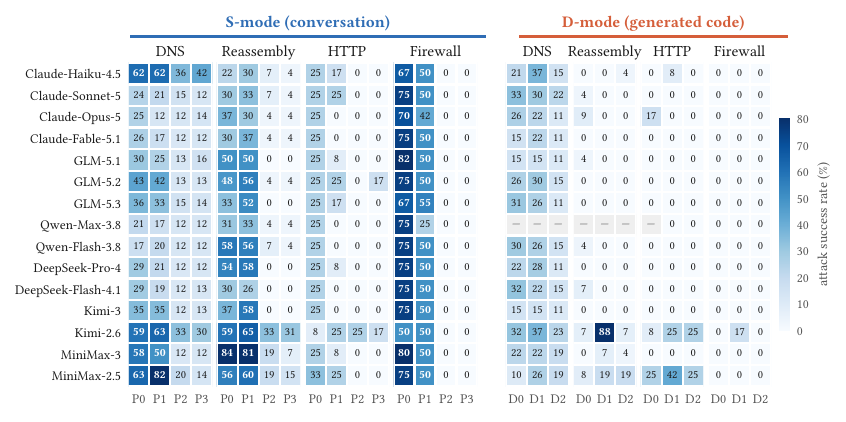}
  \caption{Attack success rate (\%) for all 15 models, per protocol, in
  S-mode (P0--P3) and D-mode (D0--D2). Darker is more
  exploitable, and ``--'' marks cells where no program was produced
  (\S\ref{sec:rq4}). Each cell pools three repetitions.}
  \label{fig:heatmap}
\end{figure}

We begin with each mode on its own, because a cross-mode comparison is only
meaningful if both modes respond to the same documents in an interpretable
way. RQ1 and RQ2 run the full matrix of 15 models $\times$ 4 protocols
$\times$ 4 S-mode levels (P0--P3) or 3 D-mode levels (D0--D2) $\times$ 3
repetitions, each repetition in a fresh session. Every S-mode conversation
and every generated program faces all V cases of its protocol
(\S\ref{sec:methodology}). Figure~\ref{fig:heatmap} reports the ASR of every
cell in both modes, and we read its left half in this section.

\noindent\textbf{DNS does not reach zero.} DNS ASR falls from 17--63\% at P0 to
12--42\% at P3 but never reaches 0\%, the floor the other protocols reach. We call the ASR
that remains once the security specification is given the \emph{residual ASR}. The
maximum belongs to Claude-Haiku-4.5 (42\%), followed by Kimi-2.6 (30\%),
the two models examined in \S\ref{sec:rq4}. The decline is also not
monotonic, since for three models (Qwen-Flash-3.8, Kimi-2.6, and MiniMax-2.5) the protocol documentation added at P1
\emph{raises} DNS ASR, for example from 63\% to 82\% for MiniMax-2.5. We read the pattern as a property
of the protocol rather than of the model.

\noindent\textbf{Reassembly behaves like the stateless protocols.}
Its ASR runs from 22\% to 84\% at P0, rises at P1 to a median of 52\% (from 37\%), and then collapses at P2 to a
median of 4\%, where five models reach 0\% and stay there at P3. Only
Kimi-2.6 (31\%) and MiniMax-2.5 (15\%) keep a residual ASR at P3:
MiniMax-3 sits at 7\%, seven models at 4\%, a single case out of 27, and
five at 0\%. Statefulness alone therefore does not predict which protocols keep a
residual ASR, and in \S\ref{sec:rq3} we ask what does.

\noindent\textbf{Stateless protocols reach zero.} HTTP S-mode ASR starts at
8--33\% at P0 and reaches 0\% at P3 for 13 models; the two exceptions are GLM-5.2 and Kimi-2.6, both at 17\% (2/12,
Wilson 95\% CI 5--45\%). For the firewall, ASR ranges from 50\% to 82\% at
P0 and falls to 0\% at P2 and P3 for all 15 models. Once the relevant rules are stated, 13 of 15 models reach 0\% on HTTP and all 15 on the firewall, so the residual risk in an unguided deployment
is a documentation problem.

\noindent\textbf{Functional documentation alone does not remove the attacks.} Both stateless
protocols reach the floor at P2, where the model first receives the security
specification, and not at P1, which adds only the functional one. At P1, no model
reaches 0\% on the firewall (25--55\%, against 50--82\% at P0) and only six do on
HTTP (0--25\%, against 8--33\%), whereas 15 and 14 models do at P2. We find that
knowing what a component must \emph{do} does not tell a model what it must
\emph{refuse}, so a conversation needs the security requirements stated.

\begin{finding}{Answer to RQ1}
\textit{(i)}~Stating the security requirements removes most attacks on firewall, HTTP, and
reassembly (median 0--4\% at P3); functional documentation alone does not.
\textit{(ii)}~DNS keeps a residual ASR of 12--42\% for every model.
\textit{(iii)}~Reassembly is stateful too but reaches the floor, so statefulness does not
explain the DNS residual ASR.
\end{finding}

\section{RQ2: Does the Same Pattern Hold When Models Write Code?}
\label{sec:rq2}

In RQ1 we found that context removes the stateless attacks from conversations
but leaves DNS with a residual ASR. We now ask whether writing code changes this
picture. The right half of Figure~\ref{fig:heatmap} shows the same
asymmetry in D-mode. At D2, DNS ASR stays between 11\% and 23\% for every model that produced
code, although D2 supplies both test feedback and the full security
specification. Reassembly, the other stateful protocol, lands with the
stateless ones rather than with DNS: 10 of 14 models reach 0\% at D2, which
rules out statefulness as the explanation of the residual DNS ASR
(\S\ref{sec:rq3}). The four exceptions are MiniMax-2.5 (19\%), Kimi-2.6
(7\%), Claude-Haiku-4.5 (4\%), and MiniMax-3 (4\%). HTTP ASR is 0\% for all
but Kimi-2.6 and MiniMax-2.5 (25\% each), and firewall ASR is 0\% for every
model. Table~\ref{tab:asr} summarizes both modes.

\begin{table}[t]
\caption{Attack success rate (\%) per protocol and level: median over models, with the
range in parentheses. S-mode levels are P0--P3 and D-mode levels are D0--D2.}
\label{tab:asr}
\small
\begin{tabular}{@{}l@{\hspace{6pt}}ccccccc@{}}
\toprule
 & \multicolumn{4}{c}{S-mode} & \multicolumn{3}{c}{D-mode} \\
\cmidrule(lr){2-5}\cmidrule(lr){6-8}
Protocol & P0 & P1 & P2 & P3 & D0 & D1 & D2 \\
\midrule
DNS & 30 (17--63) & 25 (12--82) & 12 (12--36) & 13 (12--42) & 24 (10--33) & 26 (15--37) & 15 (11--23) \\
Reassembly & 37 (22--84) & 52 (26--81) & 4 (0--33) & 4 (0--31) & 2 (0--9) & 0 (0--88) & 0 (0--19) \\
HTTP & 25 (8--33) & 8 (0--25) & 0 (0--25) & 0 (0--17) & 0 (0--25) & 0 (0--42) & 0 (0--25) \\
Firewall & 75 (50--82) & 50 (25--55) & 0 (0--0) & 0 (0--0) & 0 (0--0) & 0 (0--17) & 0 (0--0) \\
\bottomrule
\end{tabular}
\end{table}

\subsection{Security of the Programs}

\noindent\textbf{A program can pass every test and still be unsafe.} We see this
within one protocol in Kimi-2.6. Its three D1 reassembly programs each
pass the full public suite, yet 8 of 9, 8 of 9, and 6 of 7 conclusive
attacks succeed against them. The same model at D2 is attacked 0, 2, and 0
times out of 9. Reading the code shows the difference plainly. The D1
programs sort fragments by offset and concatenate them, with no mention of
overlap anywhere in the file, although the protocol note already lists the overlap
rule. The D2 programs contain an explicit \texttt{\_check\_overlap} routine and
drop the datagram when it fires. The
public tests reward a correct reassembly, not a refusal, so a green suite says nothing about the attack. CWEval likewise finds
functional but insecure code in LLM output \citep{peng2025cweval}.

\noindent\textbf{Injection through protocol data rarely succeeds.} The I cases place
instruction-like text in a TXT record or a domain label, as plain text,
Base64-encoded, or posing as an administrator. They succeed in 3 of 698 conclusive
S-mode runs and 4 of 477 D-mode runs, and never on the benign control. The four
D-mode successes are all I-DNS-005 at D0 or D1, where the program returned cached
attacker-planted data. That is cache poisoning, not obedience to an instruction, so
we leave injection out of the cross-mode comparison.

\noindent\textbf{When test feedback makes a program less secure.} Feedback does not help
every model. MiniMax-2.5's HTTP ASR is 25\% at D0, 42\%
at D1, and 25\% at D2: at D1 it passes all its functional HTTP tests, yet the
smuggling cases exploit it more often than when it had no tests. When the agent repairs the
code to pass tests on the header-rewriting paths, it can introduce a new weakness
on those paths, because the test loop optimizes for passing the tests, not for
security.

\subsection{Development Effort and Functional Correctness}

\begin{table}[t]
\caption{D-mode generation per model over the primary matrix (4 protocols
$\times$ 3 levels $\times$ 3 repetitions = 36 cells). Programs counts the
cells that produced a program, and medians are over those programs. Total
includes failed attempts. The last column repeats the DNS ASR at D2 from
Figure~\ref{fig:heatmap}.}
\label{tab:dstats}
\small
\begin{tabular}{@{}lrrrrrr@{}}
\toprule
Model & Programs & Lines & Time (s) & Cost (\$) & Total (\$) & DNS ASR D2 \\
\midrule
Claude-Haiku-4.5   & 36/36 & 226 & 382 & 0.07 &  3.60 & 15\% \\
Claude-Sonnet-5    & 36/36 & 211 & 242 & 0.32 & 13.76 & 22\% \\
Claude-Opus-5      & 36/36 & 256 & 381 & 0.70 & 25.86 & 11\% \\
Claude-Fable-5.1   & 36/36 & 211 & 323 & 1.19 & 45.38 & 11\% \\
GLM-5.1            & 36/36 & 230 & 148 & 0.83 & 31.26 & 11\% \\
GLM-5.2            & 36/36 & 220 & 410 & 0.79 & 36.84 & 15\% \\
GLM-5.3            & 36/36 & 218 & 355 & 0.73 & 35.33 & 11\% \\
Qwen-Max-3.8       & 11/36 & 249 & 430 & 0.44 &  4.81 & --  \\
Qwen-Flash-3.8     & 36/36 & 220 & 416 & 0.71 & 29.88 & 15\% \\
DeepSeek-Pro-4     & 34/36 & 192 & 272 & 0.64 & 21.84 & 11\% \\
DeepSeek-Flash-4.1 & 36/36 & 235 & 178 & 0.42 & 18.39 & 15\% \\
Kimi-3             & 36/36 & 244 & 799 & 0.80 & 27.98 & 11\% \\
Kimi-2.6           & 36/36 & 195 &  66 & 0.12 &  7.01 & 23\% \\
MiniMax-3          & 36/36 & 252 & 146 & 0.82 & 29.66 & 19\% \\
MiniMax-2.5        & 36/36 & 254 &  72 & 0.17 &  9.24 & 19\% \\
\midrule
All                & 513/540 & 224 & 258 & 0.54 & 340.85 & \\
\bottomrule
\end{tabular}
\end{table}

\noindent\textbf{What development looked like.} Table~\ref{tab:dstats} summarizes
generation per model: a median program has 224 lines of Python, and an attempt
takes 258\,s and costs \$0.54, so the primary matrix cost \$340.85. Every D1 and D2
program passed its public tests in the first harness round, so the residual
ASR is not caused by programs that fail their functional tests.

\noindent\textbf{The cost of each mode.} Per call, a conversation is lighter than a
build, as an S-mode case takes a median of three turns, 20\,s, and \$0.11, against
258\,s and \$0.54 for a D-mode program. But a program is generated once and
attacked by every case in a local sandbox, whereas S-mode needs one conversation
per case. Evaluating one model on one protocol at its strongest setting costs a
median of \$3.00 and 7.7 minutes in S-mode, against \$1.65 and 16.5 minutes for
three D2 programs. S-mode is cheaper in dollars on only 21 of the 57 such cells.
It wins on HTTP (\$0.17 against \$1.76) and loses on DNS (\$8.37 against
\$2.92) and on reassembly (\$19.29 against \$2.52). S-mode therefore saves time and the sandbox, not money, so it matters
whether its answer predicts the program's (\S\ref{sec:rq4}). The probe costs a
median of 59\,s and \$0.06 per answer.

\noindent\textbf{The programs work.} Security results are meaningful only for code
that functions. By D2, every model's
DNS program passes at least 88\% of functional cases. At D0, without tests, four models stumble on
DNS, namely Kimi-2.6 (54\%), Claude-Haiku-4.5 (62\%), Claude-Opus-5
(67\%), and Claude-Sonnet-5 (71\%). MiniMax-2.5 is worse still and passes none of its 24 DNS
functional runs. Its three D0 resolvers are complete programs, not
truncated output, since each reads a field of one dataclass from an instance of
another, so every resolution ends on that line. One test round removes it (100\% at D1). Reassembly
shows the same shape more mildly, at 92\% overall at D0 with MiniMax-2.5
(67\%) and two Claude models (73\%) lowest. Every D1 and D2 reassembly
program passes its whole suite, and so do the HTTP and firewall programs at D2,
so the results concern working code.

\begin{finding}{Answer to RQ2}
\textit{(i)}~At D2 the median model is at 0\% on firewall, HTTP, and reassembly and at 15\%
on DNS, the same pattern as in S-mode.
\textit{(ii)}~Programs that pass every public test stay exploitable, so tests that check
function do not check security.
\end{finding}

\section{RQ3: Why a Known Defense Is Missing}
\label{sec:rq3}

\begin{figure}[t]
  \centering
  \includegraphics[width=0.85\linewidth, trim=0.254cm 0cm 0.296cm 0.254cm, clip]{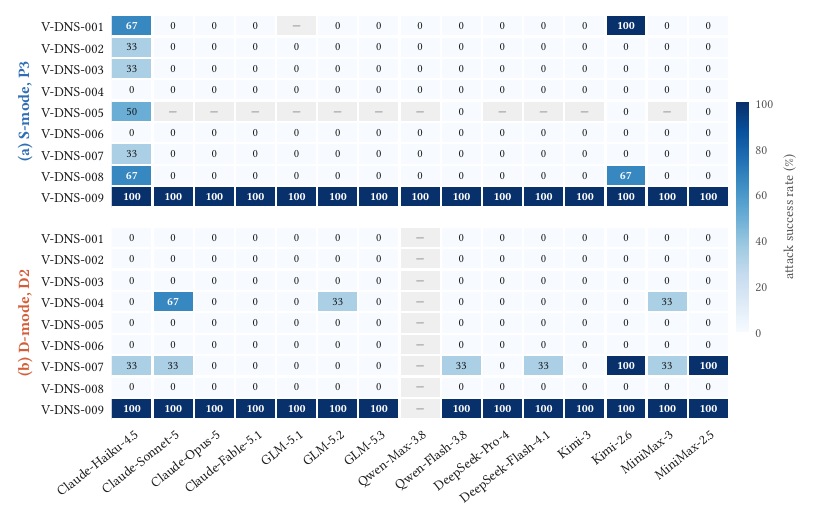}
  \caption{DNS attack success rate (\%) per V case (rows) and model
  (columns) in (a) S-mode at P3 and (b) D-mode at D2. Rows follow the case numbers of Table~\ref{tab:vcases}, and ``--'' marks cells
  without a verdict or a program.}
  \label{fig:case-heatmap}
\end{figure}

Both modes leave DNS with a residual ASR even at their richest level, while
reassembly, the other stateful protocol, reaches the floor
(\S\ref{sec:rq1}, \S\ref{sec:rq2}), so statefulness does not explain it. We
first ask whether the models know the checks that their programs lack, and then change the
specification to see what moves.

\subsection{Knowledge and the Residual ASR}

\noindent\textbf{The models know the checks.} We put one open question to each
model, with no specification, no test, and no case in front of it. It is
about to implement this component, and we ask which security checks the
component must perform. The question names no rule, and we count a rule as known if the
answer raises it unprompted, in any wording. The models name the checks. Qwen-Max-3.8 produced no program, which leaves 14 of the 15 models. On DNS,
13 of them name 0x20 validation in at least one of three answers and all 14 name
bailiwick validation; on reassembly, all 13 that answered name both the overlap rule
and the small-first-fragment rule. The one model that omits 0x20 is
MiniMax-2.5. A lack of knowledge is therefore not why these checks are missing
from the programs. Because any mention counts as naming a rule, this match may
overstate what the models know, so below we also use a stricter one, under which a model
names a check in more than half of its answers.

\noindent\textbf{Knowing does not predict the program.} To compare the probe and
S-mode on equal terms, we evaluate both against the D-mode (D2) program
(Figure~\ref{fig:ladder}). The unit is a \emph{pair}, one model and one security
check, from the five checks the probe covers (0x20, the TTL cap, and bailiwick
on DNS; overlap and small first fragment on reassembly). Each is measured for
every model with a D2 program, 14 on DNS and 13 on reassembly, which gives
$14 \times 3 + 13 \times 2 = 68$ pairs. The program lacks the check in a pair if
its D2 programs are attacked by that check's cases in at least one repetition,
the rule behind Figure~\ref{fig:case-heatmap}. That holds in 26 pairs (14 for
0x20, 7 for the TTL cap, 3 for bailiwick, 2 for the small first fragment), and
the other 42 are safe. The probe holds a check if the model names it in more than
half of its three answers, and S-mode holds it if no P3 conversation is
attacked.

In the 26 pairs where the program lacks the check, the probe's model
names it in 22 and S-mode holds it in 10; in the 42 pairs where the program has
it, the probe names it in all 42 and S-mode holds it in 41. On 0x20, S-mode
follows the program because both read the same specification, which omits the
check (below); on the TTL cap, it holds the check in 6 of 7 pairs because the
conversation applies the stated ceiling to every answer it reads. As a
screening test for vulnerable programs, the probe flags 4 of the 26 and S-mode
16 (Fisher $p = 0.0014$), and the probe flags at most 5 under any threshold we
tried. Eleven models name 0x20 in most of their answers, and all eleven write
programs that fall to it. We conclude that knowing a rule does not predict
implementing it.

\noindent\textbf{The remaining attacks concentrate in two checks.} The lower panel of
Figure~\ref{fig:case-heatmap} breaks the D2 DNS residual ASR down by case. Two
rows dominate. V-DNS-009 (0x20 case validation) is exploited in every model
that produced code, and V-DNS-007 (TTL cap) in seven. Across the three
D-mode levels and all models, these two checks account for 84\% (194/231)
of successful attacks. This concentration has two possible explanations. The
specification may not ask for these checks (a \emph{coverage gap}), or
models may be unable to implement them even when asked (a \emph{capability
limit}). The specification lists TTL capping but never mentions 0x20, apart from a
one-line field note in the protocol document. We
therefore change it in three targeted ways and measure the effect.

\subsection{Experiment 1: Adding a Missing Requirement}

We added one paragraph to the D2 security specification, asking the
resolver to randomize the letter case of every outgoing query name and to
drop any response whose echoed name does not match exactly. Nothing else
changed. If the residual ASR is a coverage gap, V-DNS-009 should drop while the
already-covered V-DNS-007 should not; if it is a capability limit, neither
should change. We regenerated DNS/D2 under the new specification three
times for each of the 14 models that produce DNS programs.

\begin{figure}[t]
  \centering
  \includegraphics[width=0.85\linewidth, trim=0.254cm 0cm 0.296cm 0.296cm, clip]{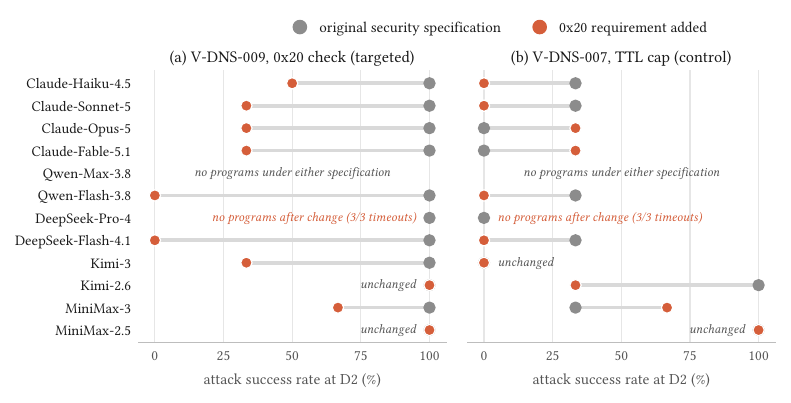}
  \caption{Per-model D2 attack success rate before (grey) and after
  (orange) adding the 0x20 requirement, for (a) the targeted check and
  (b) the control check. Each dot pools three repetitions.}
  \label{fig:intervention}
\end{figure}

\noindent\textbf{Result.} Figure~\ref{fig:intervention} shows the per-model effect.
V-DNS-009 falls from 100\% (42/42, 95\% CI 92--100) to 34\% (13/38, CI
21--50), a two-sided Fisher's exact test giving $p = 1.5 \times 10^{-11}$.
The control check does not move at all, since V-DNS-007 stands at 24\% (10/41)
before and 24\% (9/38) after ($p = 1.0$). Adding the sentence therefore
changes the check it names and nothing else. The regenerated programs
contain the check that the new paragraph asks for, which no original D2
program had; every original resolver lower-cased the echoed name before
comparing, and the regenerated program compares the echoed name against the
wire form before accepting it.

The effect is not uniform. Five models go from 3/3 to 0/3 (all three GLM
models, Qwen-Flash-3.8, and DeepSeek-Flash-4.1), five more fall to 1/3 or
1/2, and MiniMax-3 falls to 2/3, but Kimi-2.6 and MiniMax-2.5 stay at 3/3.
DeepSeek-Pro-4 produced code on every original D2 attempt but times out on
all three attempts under the revised specification (\S\ref{sec:rq4}).
0x20 has been a deployed defense since 2008
\citep{kaminsky2008dns0x20,vixie2008draft0x20}, 13 of the 14 models
describe it when asked, and not one of them wrote it until the
specification said so.

\subsection{Experiment 2: Making a Stated Rule Concrete}

Bailiwick checking is already item 2 of the specification, stated in
general terms, yet V-DNS-004, a referral whose NS record claims authority
over an unrelated zone, still succeeds occasionally. A coverage gap is
ruled out here, so we test whether the \emph{wording} is the problem,
keeping the rule and adding a concrete worked example of the V-DNS-004
scenario before regenerating DNS/D2 on the four Claude models.
\emph{The result is no effect.} V-DNS-004 succeeds on 2 of 12 runs both
before and after the change ($p=1.0$), and V-DNS-009 (12/12) and V-DNS-007
(2/12) are also unchanged. We therefore read the V-DNS-004 residual ASR as a genuine
limit on implementing a correctly stated cross-record rule.

\subsection{Experiment 3: Deleting a Stated Rule}

The first two experiments add and reword. The third deletes, on the
protocol whose specification is complete, since the reassembly specification states
seven rules that cover all nine attacks, which is why D2 reaches the floor
there (\S\ref{sec:rq2}). Each of two variants removes one rule and renumbers
the rest, either the overlap rule \citep{rfc5722} or the rule requiring the first fragment to carry the transport header \citep{rfc1858}, and drops that attack's
name from the preamble so the rule is not hinted at. We regenerated
D2 three times for each of 14 models under each variant, obtaining 39
programs per variant.

\noindent\textbf{Result: deletion is not local.} Removing the overlap rule raises
its own two cases from 0/84 to 4/76 ($p = 0.049$), and removing the
first-fragment rule raises its case from 3/42 to 6/39 ($p = 0.30$). The
cross terms are the surprise. Removing the overlap rule raises the
first-fragment case just as much, from 3/42 to 6/39, and removing the
first-fragment rule also raises the overlap cases, from 0/84 to 2/78.
Pooled over all nine V cases, attack success rises from 2\% (9/378) under
the full specification to 6\% (21/349) when the overlap rule is deleted
($p=0.015$) and 5\% (18/351) when the first-fragment rule is deleted
($p=0.076$). We find that deleting one rule degrades the program broadly, not only on
the check it named.

The asymmetry is worth stating plainly, since we cannot yet explain it.
Adding a requirement changed only the check it named, while deleting one
changed the whole program. We report that as an observation
and leave its mechanism open (\S\ref{sec:discussion}).

\begin{finding}{Answer to RQ3}
\textit{(i)}~Models name checks that their programs omit, so the probe does not predict the
program.
\textit{(ii)}~Adding a missing requirement changes only the check it names (100\% to 34\%),
but deleting a stated rule weakens the program on other checks too.
\textit{(iii)}~A worked example does not help a rule that is already stated.
\end{finding}

\section{RQ4: How Far Does S-Mode Track the D-Mode Program?}
\label{sec:rq4}

RQ1 and RQ2 showed that, across all models, the two modes follow the same
pattern, and RQ3 traced the residual DNS ASR to the specification. That does not
show that S-mode predicts the program that a given model writes, which decides
whether S-mode can stand in for D-mode. We compare the modes at P3 and D2, where
both hold the full security specification (\S\ref{sec:methodology}).

\subsection{S-Mode as a Screening Test}

\noindent\textbf{Screening against the D-mode program.} We ask how many real vulnerabilities
S-mode misses and how many false alarms it raises. We evaluate it against D-mode
as a screening test over all $15 \times 9$ (model, case) pairs at DNS, reducing
each pair to its majority verdict over the three repetitions and dropping pairs
with no conclusive run in either mode, which leaves 115. D-mode is the reference,
since the program is what ships. \emph{Sensitivity} is the share of vulnerable
pairs that S-mode flags, \emph{specificity} the share of safe pairs that it
leaves unflagged, and the \emph{negative predictive value} (NPV) the share of its unflagged
pairs that are safe. S-mode reaches a sensitivity of 0.824 (14 of 17), a
specificity of 0.959 (94 of 98), and an NPV of 0.969 (94 of 97). Cohen's $\kappa$, which measures agreement
beyond chance \citep{cohen1960coefficient}, is 0.764, which counts as substantial
agreement \citep{landis1977measurement}. It
raises four false alarms and misses three vulnerabilities. S-mode and D-mode
flag nearly the same number of pairs (18 and 17), but they disagree on seven,
over four models, so similar totals hide disagreement on individual pairs. The main weakness is that roughly one real DNS vulnerability in six goes
unflagged, and S-mode alone would ship those pairs. S-mode is faster than a build but not cheaper
(\S\ref{sec:rq2}), so what it buys is time and the absence of a sandbox, and the
miss rate sets its floor.

\begin{figure}[t]
  \centering
  \includegraphics[width=0.8\linewidth, trim=0.254cm 0cm 0cm 0.339cm, clip]{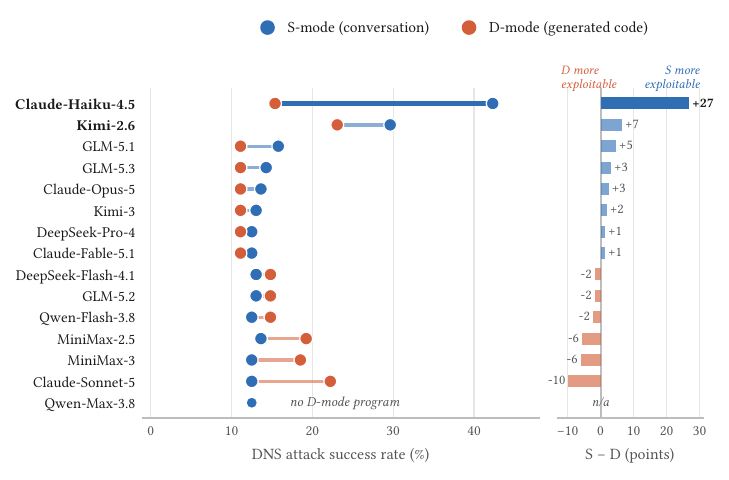}
  \caption{DNS attack success rate per model in S-mode at P3 (blue) and
  D-mode at D2 (orange), with the signed difference as a diverging bar on
  the right. Rows are ordered by the S$-$D difference, and bold labels mark
  the two models examined below.}
  \label{fig:svsd}
\end{figure}

\noindent\textbf{The two modes mislead in opposite directions.} HTTP and the
firewall carry no such risk, since the firewall is 0\% for all 15 models in both
modes, and HTTP is 0\% for 12, with the three exceptions small (17--25\%)
and one of them, Kimi-2.6, non-zero in both. DNS is where the modes come
apart, and Figure~\ref{fig:svsd} plots the pair for each model. The mean
absolute S--D difference is 5.4 points. For 12 models the signed gap lies
between $-10$ (Claude-Sonnet-5) and $+5$, with no consistent sign. Two
models stand out, and for both S-mode is \emph{more} exploitable than
D-mode. Claude-Haiku-4.5 reaches 42\% against 15\%, a 27-point gap, and
Kimi-2.6 reaches 30\% against 23\%. These gaps are large, but we cannot separate them from chance. A two-sided Fisher exact test on each model's P3 and D2 DNS
outcomes gives a smallest p-value of $0.064$ (Claude-Haiku-4.5), with every
other model at $p \geq 0.47$, so at three repetitions per cell no individual
model separates the modes on its overall rate. The comparison per case gives
stronger evidence. We say that S-mode \emph{under-reports}
when its conversation fails a check that the program has, and \emph{over-reports}
when it holds a check that the program lacks.

\subsection{Under- and Over-Reporting}

\noindent\textbf{Under-reporting: S-mode fails a check that the program has.}
S-mode under-reports on two models, Claude-Haiku-4.5 and Kimi-2.6.
Claude-Haiku-4.5 has 11 S-mode attack successes at P3, and only V-DNS-009 among
them is shared with its larger siblings, and the rest span six checks that
the security specification states explicitly. In V-DNS-001, the driver
sends a spoofed response with the wrong transaction ID, followed by the
legitimate one.
\begin{codebox}
model  -> {"ev":"upstream_query","dst":"203.0.113.1:53","msg":{"id":1001,...}}
          {"ev":"state","pending":[{"msg_id":1001,"dst":"203.0.113.1:53",...}]}
driver -> {"ev":"upstream_response","msg":{"id":1002,"answer":[..."6.6.6.6"]}}
model  -> {"ev":"client_response","msg":{"answer":[..."data":"6.6.6.6"]}}
\end{codebox}
\noindent The model answers with the attacker's address from the first,
mismatched message. Its prompt requires accepting only a response that
matches the outstanding ID, and the matching response arrives one turn
later. At P3, the model had listed the pending ID in its emitted state,
which was handed back with the spoofed message. It does not forget the ID
but fails to check the message against it. Its D2 program handles the case
correctly, because there the rule became a lookup that runs on every
message (\texttt{self.pending.get((txid, src))}), and its other S-mode
failures (V-DNS-002, -003, -005, -007, and -008) have a similar shape. Each
involves either a race between a spoofed and a legitimate message, or
earlier state, such as a negative-cache entry or a TTL ceiling, that must be
checked again later. Kimi-2.6 fails V-DNS-001 (3/3) and V-DNS-008 (2/3) in
S-mode through the same first-response-wins mechanism, while its D2 code
keys responses on the transaction ID. No other model shows this pattern.
The
two diverging models are the smaller or cheaper members of their series, but
size and price are confounded within each series, so we leave capability as a
hypothesis. We therefore do not tie the failure to one vendor.

\noindent\textbf{Over-reporting: S-mode holds a check that the program lacks.} S-mode over-reports on the TTL cap (Figure~\ref{fig:case-heatmap}). In S-mode at P3, 14 of 15 models
handle V-DNS-007 (TTL cap), yet in D-mode at D2 the programs of seven
models are exploitable. Capping a TTL needs no earlier state, so this is
not a recall problem but an omission in code. The conversational agent
applies the stated ceiling whenever it reads an answer, but the resolvers generated by seven models copy the TTL into the cache
unchanged. By design, the
public tests at D1 and D2 contain no security checks, as they verify only that
a TTL is honored on expiry, never that it is capped, so test-driven
development rewards honoring any TTL. This is why six of the 15 gaps in
Figure~\ref{fig:svsd} are negative, such as $-10$ for Claude-Sonnet-5.
We find neither mode uniformly stricter, since each misses different checks.

\subsection{What S-Mode Measures and Misses}

\noindent\textbf{S-mode measures something a program cannot.} Under-reporting is a
defect when S-mode is used to predict a program, but it also measures whether a
model keeps a rule across turns. In V-DNS-001 the model received the pending ID,
restated it in its own state, and then failed to compare against it one message
later. A program cannot show this failure, because its check either exists or
does not. The same property matters wherever the model stays in the loop and no
program is written, such as honeypots and other conversational services
\citep{sladic2024shell,fan2024honeyllm}, agents driving multi-step operations,
and long-context workflows \citep{laban2026lostmultiturn,liu2024lostmiddle}.
There D-mode has nothing to build, so we regard the two modes as complementary,
not as substitutes.

\noindent\textbf{S-mode cannot see a failure to build.} Some models produce no
program in D-mode. Qwen-Max-3.8 timed out on all nine DNS and all nine
reassembly attempts at our 1{,}800\,s budget, and on seven of nine HTTP attempts
but none of nine firewall attempts. It reads the three documents in under half a
second and then emits nothing for thirty minutes. DeepSeek-Pro-4 fails more
narrowly, on one DNS D0 repetition three times in a row and on all three attempts
under the 0x20-revised specification (\S\ref{sec:rq3}). S-mode always returns a
reply, malformed in only 16 of 9{,}360 case runs, and the S-mode DNS profile of
Qwen-Max-3.8 is unremarkable among the 15 (Figure~\ref{fig:heatmap}). S-mode
therefore cannot flag a model that fails to produce a program.

\noindent\textbf{Repeatability.} The two modes are equally repeatable, and both are least repeatable on DNS.
There, 87.9\% of S-mode and 85.7\% of D-mode (model, level, V case) triples give
the same verdict in all three repetitions, against 94\% or more elsewhere, so a
single-run evaluation would misjudge about one DNS outcome in eight.

\begin{finding}{Answer to RQ4}
\textit{(i)}~S-mode flags 82\% of the real DNS vulnerabilities and leaves 96\% of the safe
cases unflagged.
\textit{(ii)}~It over-reports checks that D-mode lacks and under-reports checks that D-mode
has.
\textit{(iii)}~S-mode can therefore screen but not replace D-mode testing.
\end{finding}

\section{Threats to Validity}
\label{sec:threats}

Our conclusions rest on automated verdicts, a fixed set of models and
protocols, and small samples. We discuss each threat and how we mitigated it.

\noindent\textbf{Internal validity.} Verdicts come from one automated checker per
security property, validated against a reference implementation (\S\ref{sec:methodology}),
not from manual review of all 9{,}360 S-mode and
8{,}661 D-mode case runs. We did review by hand every transcript and code excerpt in
this paper against its verdict. This found two checker bugs, which we fixed
before computing the reported results. Three DNS functional checkers compared
an internal query identifier that the implementation chooses freely, and the
case engine treated a legitimate strategy, coalescing duplicate queries, as an
infrastructure error. We re-ran the checkers on the 31 affected cases.  

\noindent\textbf{Construct validity.} A binary \texttt{attack\_success} verdict is a
proxy for exploitability in one scripted scenario, not a measure of real
harm. We wrote our V cases from documented attack classes (\S\ref{sec:background}) rather than
sampling them from attack traffic or producing them by adaptive search, so
a model could handle our phrasing yet fail on semantically equivalent
variants.

\noindent\textbf{Mode comparison.} For reassembly, HTTP, and the firewall, the D-mode
protocol note lists the security properties, so D0 and D1 are not free of security
requirements there, unlike S-mode at P0 and P1 (\S\ref{sec:methodology}). We
therefore compare modes only at P3 and D2, and read RQ4 as a comparison of two
realistic workflows at their strongest settings. \S\ref{sec:rq3} is the design that
isolates a single variable.

\noindent\textbf{Statistical conclusion validity.} The specification experiments
report Fisher's exact tests on 12--84 conclusive runs per condition, so they can
detect the large effect on V-DNS-009 but not a small one on V-DNS-007 or
V-DNS-004; a null result there means ``no detectable change'', not ``no
change''. We ran about ten tests across the three experiments. The 0x20 effect
($p = 1.5 \times 10^{-11}$) survives a Bonferroni correction for all of them,
whereas the deletion results ($p = 0.049$ and $0.015$) do not, so \S\ref{sec:rq3}
reports deletion as an observation, not a confirmed effect. The same caution
applies to the ASR differences of \S\ref{sec:rq4}, which we therefore rest on
case-level agreement.

\noindent\textbf{External validity.} We study network protocol implementations only, namely four protocols, two stateful and two stateless, whose behavior is a function from messages to messages (\S\ref{sec:intro}). Whether the same holds for software with other interfaces is untested. The 15 models were chosen by availability during the study.
The reassembly result does not support statefulness as the explanation of the
residual DNS ASR. With two protocols per class we can say that statefulness does
not by itself predict the residual ASR, but not what does in general. Figure~\ref{fig:ladder} has 26 vulnerable pairs from four checks, 14 of them
0x20, so we show that the probe over-reports on these checks, not on every
rule, and we matched its TTL-cap answers by keyword without checking them by
hand. The
knowledge probe also asks for a checklist, which is easier than recalling a rule
while writing code, so it may overstate what models know when they write code. 

\section{Related Work}
\label{sec:related}

Our study relates to three lines of work, namely the security of LLM-generated code,
LLMs that simulate systems, and the validity of LLM benchmarks. The first
evaluates generated code alone, the second studies simulators without checking
them against the system the same model would build, and the third questions
whether a benchmark measures what its labels claim. \textit{To our knowledge, no prior work holds one task, one attack suite, and one oracle fixed across S-mode, D-mode, and the probe.}

\noindent\textbf{Security of LLM-generated code.} Pearce et al.\ found that about
40\% of 1{,}689 Copilot completions for security-relevant scenarios were
vulnerable \citep{pearce2022asleep}, and Perry et al.\ that developers using an AI
assistant wrote less secure code while believing it more secure
\citep{perry2023insecure}. Later work shows that security prompting helps without
eliminating vulnerabilities
\citep{tony2025prompting,promptmethodsempirical2026,securingcodegen2026,illusionsecure2026},
scores prompts by weakness class \citep{cheng2024pyp4llmsec,siddiq2024sallm}, and
moves to repository-level tasks \citep{secrepobench2025}. CWEval checks
functionality and security of the same code and finds functional but insecure
code \citep{peng2025cweval}, and re-evaluating secure-code-generation methods
shows that they can trade functionality for security \citep{dai2026rethinking}.
All of this evaluates generated code on its own. To our knowledge, none asks
whether its vulnerability can be predicted from how the model behaves when asked to
act as the system. ClarifyGPT has the model ask clarifying questions about
ambiguous requirements before it codes \citep{mu2024clarifygpt}. Our probe instead
asks what a model can state about its security requirements, and we find that
stating them does not put them in the code. We also add an S-mode control on the
same cases, specification experiments that ask \emph{why} prompting falls short,
and stateful tasks attacked over many messages.

\noindent\textbf{LLMs that simulate systems and people.} Role play is a basic LLM
behavior \citep{shanahan2023roleplay}. LLMs simulate people
\citep{park2023generative}, end users' security attitudes
\citep{simulateendusers2026}, and services such as shells and HTTP servers
in honeypots \citep{sladic2024shell,fan2024honeyllm,otal2024llmhoneypot}.
Adversarial role play can override safety behavior
\citep{tang2025darkness,roleplaystress2026}, and our S-mode failures resemble
it. Other
work questions the simulator itself, since LLM-simulated users match
real behavior on only 6--8\% of style dimensions
\citep{zhu2026realusersim}, and one security benchmark keeps separate tracks
for simulated and real execution \citep{raseval2025}. To our knowledge, none of this checks simulated behavior against the same model building the simulated system, which is our comparison. Multi-turn studies report that models lose earlier
instructions and mid-context information
\citep{laban2026lostmultiturn,liu2024lostmiddle}, consistent with the
mechanism in \S\ref{sec:rq4}.

\noindent\textbf{Validity of LLM4SE benchmarks and LLMs for networks.} Existing
work also asks whether coding-agent benchmarks measure what their labels claim.
A task's category label is an unreliable proxy for what the task demands
\citep{shayanfar2026agentic}, and end-to-end resolution rates merge model,
harness, and environment into one score \citep{gorinova2026misaligned}. A manual audit of SWE-bench Lite found leaked patches and missing information \citep{xia2025agentless}. All three bear on our concern about using S-mode as a proxy, which is why we
also report case-level agreement in \S\ref{sec:rq4}. In networking, LLMs
extract protocol grammars and generate fuzzing messages
\citep{meng2024chatafl} and translate requirements into configurations
\citep{wang2024netconfeval}; these use LLMs as tools for engineers, whereas we evaluate what an LLM produces when asked to be, or to build, the
protocol implementation.

\section{Discussion}
\label{sec:discussion}

\noindent\textbf{Lessons learned.} For \emph{build reliability}, pipelines must detect a
run that produces nothing, as Qwen-Max-3.8 did for DNS. For \emph{functional versus
security correctness}, programs passed their public tests yet stayed exploitable
(\S\ref{sec:rq2}), so adversarial tests belong in the continuous-integration pipeline. For \emph{choosing between
the modes}, the probe cannot approve a model, and S-mode is a first filter with a
known floor, since about one real vulnerability in six escapes it, so the D-mode pass
should go to the checks S-mode misses. Finally, specifications should name the
checks they need (\S\ref{sec:rq3}).

\noindent\textbf{Recommendations for researchers.} We recommend four practices for
studies of LLM security. \textit{(i)}~Judge the program an agent builds, not only
its answers or conversation. \textit{(ii)}~State which requirements the
specification names. \textit{(iii)}~Repeat each run, since about one DNS outcome in
eight differs across repetitions. \textit{(iv)}~Validate the oracle against a
secure reference implementation.

\section{Ethical Considerations}
\label{sec:ethics}

We attack only artifacts built for this study, against programs in a network-isolated sandbox,
and never deploy them. The attacks encode long-documented weaknesses
(\S\ref{sec:related}), and no human subjects or private data are involved.

\noindent\textbf{Use of generative AI.} An LLM assistant was used only to polish the
language of this paper. It did not produce the study design, the experiments,
the analysis, or the claims.

\section{Conclusion}
\label{sec:conclusion}

We compared S-mode, D-mode, and a knowledge probe. The probe does not predict
the program, and S-mode is a useful filter, but only testing the built code shows
what an agent ships.


\bibliographystyle{ACM-Reference-Format}
\bibliography{reference}

\end{document}